\documentstyle[twocolumn,aps,epsfig,floats]{revtex}
\begin{document}

\twocolumn[\hsize\textwidth\columnwidth\hsize\csname %
@twocolumnfalse\endcsname

\title{Electronic Structure near Impurities in the Chains of
YBa$_2$Cu$_3$O$_{6+x}$}
\author{Dirk K. Morr and Alexander V. Balatsky}
\address{Theoretical Division, MS B262, Los Alamos National
Laboratory, Los Alamos, NM 87545}
\date{\today}
\draft \maketitle

\begin{abstract}
We study the electronic structure near impurities in the
one-dimensional chains of YBa$_2$Cu$_3$O$_{6+x}$. Assuming chain
superconductivity below $T_c$ due to a proximity coupling, we show
that only a magnetic impurity induces resonance states in the
local density of states. Its spatial structure reflects the
particle-hole nature of chain superconductivity and thus
distinguishes it from other broken symmetry phases. For a
sufficiently large scattering strength, or due to interference
effects arising from a second impurity, the chains undergo a
quantum phase transition to a polarized state.
\end{abstract}


]

Studying the electronic structure of the CuO$_2$ planes in
YBa$_2$Cu$_3$O$_{6+x}$ (YBCO), an important step towards our
understanding of superconductivity, has been complicated by the
presence of one dimensional (1D) CuO-chains in this material. The
coupling of the CuO-chains to the CuO$_2$-planes changes the
electronic structure of the latter and might facilitate the
occurrence of other 1D structures, e.g., stripes, in the planes.
Much theoretical work \cite{theory1,theory2} has focused on this
coupling and its implications for a series of experimental probes
ranging from inelastic neutron scattering \cite{Mook}, nuclear
quadrupole resonance (NQR) \cite{Gre00,NQR}, infrared spectroscopy
\cite{Bas95}, angle-resolved photoemission \cite{Sch98} and
scanning tunneling microscopy (STM) experiments \cite{Edw94}. In
turn, the same coupling necessarily changes the chain electronic
structure when the CuO$_2$ planes become superconducting (SC), as
was recently evidenced by NQR \cite{Gre00,NQR} and STM
\cite{Der01} experiments. In particular, NQR experiments observe
an increase in the $^{63}$Cu NQR line width near the SC transition
\cite{Gre00}, while STM experiments \cite{Der01} find evidence for
large oscillations in the local density-of-states (DOS). Whether
these observations arise from the presence of a charge-density
wave (CDW) \cite{Gre00}, or from impurities in SC chains is
currently a topic of intense discussion.

The study of the electronic structure near impurities in 2D
superconductors \cite{th2D,exp2D} has been a successful method to
distinguish between various broken symmetry ground states. In this
Letter we extend these works to study the effects of chain
impurities, both magnetic and non-magnetic, on the local DOS
assuming that the chains become SC due to a proximity coupling.
Both types of impurities induce large spatial oscillations in the
DOS, and thus the electron density, whose amplitude increases
below $T_c$. However, in contrast to a non-magnetic impurity, a
magnetic impurity induces two resonance states in the chain DOS.
Moreover, we present a prediction for the spatial dependence of
these resonances which reflects the particle-hole nature of chain
superconductivity. This prediction therefore discriminates between
oscillations arising from a CDW and those due to impurities.
Finally, we find that as the distance between two magnetic
impurities is reduced, the resonance states split and the system
can be driven through a quantum phase transition. Our work is
complementary to that of Eggert \cite{Egg00} who studied the
effects of impurities in a 1D Luttinger liquid.

Since in the following we employ a model described in detail in
Ref.~\cite{Morr00a}, we only briefly review it here. The
mean-field BCS-Hamiltonian of the chain-plane system is given by
\cite{theory1,Morr00a}
\begin{eqnarray}
{\cal H} &=& - \sum_{k, \sigma} \epsilon_{\bf k}
c^\dagger_{k,\sigma} c_{k,\sigma} - \sum_{k, \sigma} \zeta_{\bf k}
d^\dagger_{k,\sigma} d_{k,\sigma} - t_\perp \sum_{k, \sigma}
c^\dagger_{k,\sigma} d_{k,\sigma} \nonumber \\ & & \quad + \sum_k
\Delta_k c^\dagger_{k,\uparrow} c^\dagger_{k,\downarrow}  + h.c.
\label{Hpc}
\end{eqnarray}
where $c^\dagger_k, d^\dagger_{k_x}$ are the fermionic creation
operators in the plane and chain, respectively. One has for the
planar and chain tight-binding dispersions, $\epsilon_{\bf k}$ and
$ \zeta_{\bf k}$,
\begin{eqnarray}
\epsilon_{\bf k} &=& -2t_p \Big( \cos(k_x) + \cos(k_y) \Big)
 -4t_p^\prime \cos(k_x) \cos(k_y)  -\mu_p \ , \nonumber \\
\zeta_{\bf k} &=& -2 t_c \cos(k_x) -\mu_c \ .
\label{dispersion}
\end{eqnarray}
Here, $t_p=300$ meV, $t^\prime_p=-0.22t_p , \mu_p= -0.812t_p $ are
the planar hopping elements and the chemical potential
\cite{Sch98}, and $\Delta_{\bf k}=\Delta_0 \ (\cos(k_x) -
\cos(k_y))/2$ is the planar SC gap with $\Delta_{0} \approx 40$
meV \cite{Mag96}.  The chain Fermi momentum, $k^c_F \approx 0.25
\pi$, \cite{Mook,Edw94} yields $\mu_c=-1.41t_c$, and we take
$t_\perp=0.4t_p$ \cite{Morr00a}. After integrating out the planar
fermions, the chain Greens function $\hat{G}_0 =- \langle {\cal T}
\Psi_{k_x,l}(\tau) \Psi^\dagger_{k_x,m}(\tau^\prime) \rangle$ with
\begin{equation}
\Psi_{k_x,l}^\dagger=(d_{k_x,l,\uparrow}^\dagger,
d_{k_x,l,\downarrow}^\dagger, d_{-k_x,l, \uparrow}, d_{-k_x,l,
\downarrow}) \ ,
\end{equation}
and $m,l$ chain indices, is given by
\begin{equation}
{\hat G}_0^{-1}= \left(
\begin{array}{cc}
B(k_x,\omega_n) \, \hat{\sigma}_0 & C(k_x,\omega_n) \,
\hat{\sigma}_0
\\ C(k_x,\omega_n) \, \hat{\sigma}_0 & - B(-k_x,-\omega_n)
\, \hat{\sigma}_0
\end{array}
\right)   \, , \label{Ginv}
\end{equation}
where $\hat{\sigma}_0$ is the unit matrix in spin space, and
\begin{eqnarray}
B &=& \ \left( \, i\omega_n - \zeta_{\bf k} \, \right) \
\delta_{l,m} \nonumber \\ & & \hspace{-0.5cm} -t_\perp^2 \, N^{-1}
\sum_{k_y}  \,  {i \omega_n + \epsilon_{\bf k} \over (i
\omega_n)^2 - (\epsilon_{\bf k})^2 - \Delta_k^2 } \
 e^{ ik_y(l-m) }  \, ; \nonumber  \\
C &=& -t_\perp^2 \, N^{-1}  \sum_{k_y}
 {\Delta_k \over (i \omega_n)^2 -
(\epsilon_{\bf k})^2 - \Delta_k^2 } \ e^{ ik_y(l-m) } \, .
\label{G0F0}
\end{eqnarray}
In contrast to previous work \cite{theory1}, we showed
\cite{Morr00a} that the experimentally measured clean DOS can only
be obtained {\it qualitatively} as well as {\it quantitatively} by
making the ansatz
\begin{equation}
{\hat G}_c^{-1}(k_x,l,m,\omega_n) = {\hat G}_c^{-1}(k_x,\omega_n)
\ \delta_{l,m} \, . \label{ans}
\end{equation}
We argued that the physical origin of Eq.(\ref{ans}),
which describes the absence of {\it coherent} correlations between
the chains, or equivalently, the coupling of a single chain, lies
in the presence of planar stripes and/or doping inhomogeneities in
the chains.

\begin{figure} [t]
\begin{center}
\leavevmode
\epsfxsize=7.5cm
\epsffile{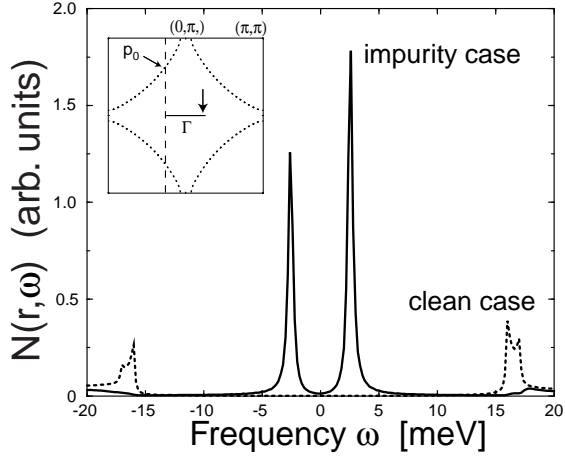}
\end{center}
\caption{Chain DOS in the SC state for the clean case (dashed line)
and at the site of a magnetic impurity with
$\beta=200$ meV and $U_0=0$ (solid
line). We set the lattice
constant $a_0=1$. Inset: Fermi sea of the chain
(solid line) and
Fermi surface of the plane (dotted line). }
\label{DOSmag}
\end{figure}
We employ the $\hat{T}$-matrix formalism \cite{Shi68} to
obtain the local DOS
\begin{equation}
N(\omega)=-{2 \over \pi} \ {\rm Im} \left[ \hat{G}^{(11)}({\bf
r,r},\omega_+)+\hat{G}^{(22)}({\bf r,r},\omega_+) \right] \ ,
\end{equation}
in the vicinity of an impurity, with $\omega_+=\omega+i0^+$. Here,
one has for the full chain Greens function
\begin{eqnarray}
\hat{G}({\bf r},{\bf r}',i \omega_n) &=& \hat{G}_0({\bf r}-{\bf
r}^\prime,i\omega_n) + \nonumber \\ & & \hspace{-1.0cm}
\hat{G}_0({\bf r},i\omega_n) \hat{T}(i\omega_n) \hat{G}_0(-{\bf
r}',i\omega_n) \ , \label{fullG}
\end{eqnarray}
where $\hat{T}(i\omega_n)$ represents all scattering processes and
$\hat{G}_0$ is obtained from Eqs.(\ref{Ginv}) and (\ref{ans}). In
what follows we consider quite generally an impurity which is both
a magnetic and potential scatterer. Assuming for simplicity that
the spin of the impurity is aligned along the $z$-axis \cite{com1}
we obtain for the scattering Hamiltonian
\begin{equation}
{\cal H}_{sc}=\sum_{k_x, k_x^\prime, m} \Psi_{k_x,m}^\dagger
\hat{V} \Psi_{k_x^\prime,m} \label{Hnmimp}
\end{equation}
with $\hat{V}=\hat{\tau}_3 \left( U_0 \hat{\sigma}_0-JS_z
\hat{\sigma}_3 \right)$ and $\tau_i,\sigma_i$ are the
Pauli-matrices in Nambu and spin-space, respectively. $U_0 \, (J)$
is the potential (magnetic) scattering strength and $S_z$ is the
$z$-projection of the impurity spin.  One then finds
\begin{equation}
\hat{T}(i\omega_n)= \left[ 1- \hat{V} \hat{G}_0 (r_x=0,i\omega_n)
\right]^{-1} \, \hat{V}  \ .
\label{T}
\end{equation}
We remind that an impurity bound state exists at a frequency
$\omega_{res}$, if $\det[\hat{T}^{-1}(\omega_{res})] \equiv 0$, while for a
resonance state only the real part of
$\det[\hat{T}^{-1}]$ vanishes.

We first study the case of a purely magnetic impurity with $U_0=0$
and $J \not = 0$. A straightforward analysis of Eq.(\ref{T}) shows
that there exist two resonance states whose frequencies,$\pm
\omega_{res}$, are given by the solution of
\begin{equation}
{ { \omega_{res} \over \alpha {\bar \Delta} } } \left[\sqrt{ {\bar
\Delta}^2 - \omega_{res}^2}+  \alpha \right] = \pm { \gamma^2 -1
\over \gamma^2 + 1}
\end{equation}
where $\gamma= v^c_F/\beta$, $v^c_F$ is the chain Fermi velocity
and $\beta=JS_z/2$. Moreover, $\alpha= t_\perp^2 / {\bar v}$,
${\bar v} = \left( \partial \epsilon_{\bf k} / \partial k_y
\right)|_{{\bf p}_0} $,  where ${\bf p}_0$ is the momentum on the
planar Fermi surface (FS) with $p_0^x=k^c_F$ (see inset of
Fig.~\ref{DOSmag}) and ${\bar \Delta}=\Delta({\bf p}_0)$. While in
the limit $\gamma\ll 1$, the resonance frequencies are located
close to $\pm {\bar \Delta}$, $\omega_{res}$ shifts to lower
energies with increasing $\gamma$. At the critical $\gamma_c=1$
(i.e, $\beta_c=v_F^c$) the resonance states cross at
$\omega_{res}=0$ and the magnetic impurity forms a bound state
with a spin-up (down) quasi-particle for $J<0 \, (>0)$.
Accordingly, the ground state of the system undergoes a quantum
phase transition and changes from an unpolarized one for
$\gamma<\gamma_c$ to a spin-polarized one with total spin $S=1/2$
for $\gamma>\gamma_c$. A similar transition also occurs in
superconductors with $s$-wave symmetry \cite{swave,Balatsky}.

\begin{figure} [t]
\begin{center}
\leavevmode
\epsfxsize=7.5cm
\epsffile{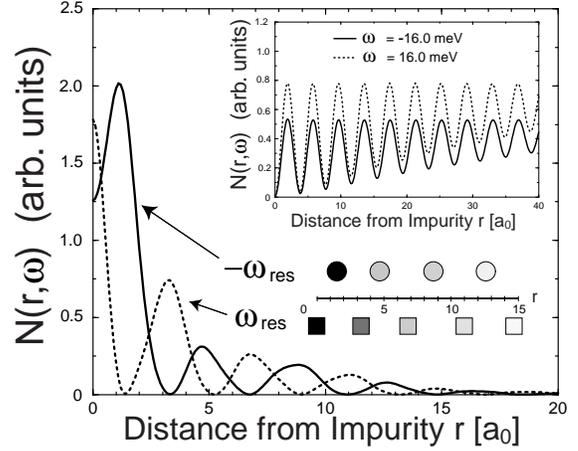}
\end{center}
\caption{Chain DOS as a function of distance from the impurity
site for the resonance states at $\omega_{res}=\pm 2.6$ meV and
(see upper inset) for the coherence peaks at the gap edges,
$\omega_{cp}=\pm 16.0$ meV. A gray-scale intensity plot of the DOS
at $\pm \omega_{res}$ is shown in the lower right.}
\label{DOSosc}
\end{figure}
In Fig.~\ref{DOSmag} we present the SC chain DOS, obtained from a
numerical evaluation of Eq.(\ref{fullG}), in the clean case and at
the site of a magnetic impurity with $\beta=200$ meV and $U_0=0$.
A comparison with the clean DOS shows the emergence of two
resonance  states at the impurity site with $\omega_{res} \approx
\pm 2.6$ meV, in agreement with the discussion above.
Simultaneously, the impurity suppresses the coherence peaks at
$\omega_{cp}= \pm 16.0$ meV which are present in the clean DOS.
Moreover, the impurity induces spatial oscillations in the chain
DOS, which we present in Fig.~\ref{DOSosc} for the two resonance
states and the coherence peaks at the gap edges (see inset). The
oscillations at $\pm \omega_{res}$ possess a broad range of
wave-vectors with $0 \leq k_{osc} \leq 1.45$, which gives rise to
the rapid decay of the oscillations' amplitude with distance from
the impurity. The decay length, $\xi \approx 10-15a_0$ is
consistent with the experimentally obtained value of $40\AA$
\cite{Der01}. Note that the oscillations at $\pm \omega_{res}$ are
out-of-phase, such that a maximum in the DOS for $+\omega_{res}$
coincides with a minimum for $-\omega_{res}$, as is particularly
apparent when one plots the spectral weight of the resonance
states at their peak positions (see inset (b)). The complementary
intensity pattern in real space of the hole-like and particle-like
resonance states is a reflection of the particle-hole nature of
Bogoliubov quasi-particles \cite{exp2D,ph}. This characteristic
feature of chain superconductivity is {\it qualitatively}
different from the response of a CDW, and hence  discriminates
between proposed broken symmetry phases in the chains. In
contrast, the oscillations at $\pm \omega_{cp}$, the frequency of
the coherence peaks (see inset), are in-phase. Moreover, their
amplitude decays much slower with distance from the impurity site
since the wave-vectors of these oscillations are centered in a
narrow range, $\Delta k \approx 0.02$, around $k_{osc}=0.81$ (see
arrow in the inset of Fig.~\ref{DOSmag}). The wavelength,
$\lambda$, of these oscillations exhibits an asymmetric frequency
dependence with $\lambda(-\omega_0)>\lambda(\omega_0)$ for
$\omega_0>0$ and a minimum at $\omega \approx 0$, in agreement
with STM experiments \cite{Der01}. The spatial dependence of the
DOS oscillations is qualitatively robust against changes in the
form of the planar FS.  However, their amplitude increases in the
SC state, and accordingly, the spatial electron density changes.
While this result is consistent with the broadening of the NQR
line in the SC state \cite{Gre00,NQR}, a quantitative analysis
requires a microscopic model for the coupling of the electron
density to the quadrupolar moment which is beyond the scope of
this Letter.

We next consider a purely non-magnetic impurity with $J=0$ and
$U_0 \not = 0$. We remind that a potential scatterer only induces
a resonance state if the SC phase varies along the FS, or between
Fermi points. However, the two $k_F$-points in the chain possess
the {\it same} SC phase, which leads to $\det[\hat{T}^{-1}] \sim
1+(U_0/v^c_F)^2$, and no resonance states exist, in analogy to the
$s$-wave superconductor. Our numerical evaluation supports this
conclusion revealing no resonances but again large spatial
oscillations in the DOS.

\begin{figure} [t]
\begin{center}
\leavevmode
\epsfxsize=7.5cm
\epsffile{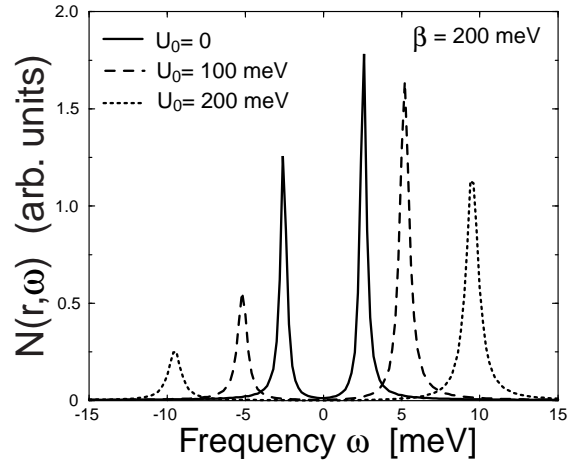}
\end{center}
\caption{Frequency dependence of the chain DOS at the impurity site for
$\beta = 200$ meV and $ U_0 = 0$ (solid line), $U_0 = 100$ meV
(dashed line), and $U_0 = 200$ meV (dotted line).} \label{DOSbU}
\end{figure}

Since an impurity is in general a magnetic and potential scatterer
we next study the case $J, U_0 \not =0$ and present in
Fig.~\ref{DOSbU} the chain DOS at the impurity site for $\beta =
200$ meV and three values of $U_0$.
We find that with increasing $U_0$, the position of the resonances
shifts up in energy from $\omega_{res}=2.6$ meV for $U_0=0$ to
$\omega_{res}=9.5$ meV for $U_0=200$ meV. This implies, that the
critical value $\gamma_c$ (and thus $\beta_c$) where the ground
state of the system becomes spin-polarized increases, again
similar to the case of an $s$-wave superconductor \cite{Balatsky}.
Moreover, while the spatially integrated spectral weight of the
particle-like resonance is $1.43$ times larger than that of the
hole-like resonance for $U_0=0$, the ratio is $0.77$ for $U_0=200$
meV. Thus spectral weight is transferred from the particle-like to
the hole-like resonance with increasing $U_0$, however, their
phase-shifts remain unchanged.

\begin{figure} [t]
\begin{center}
\leavevmode
\epsfxsize=7.5cm
\epsffile{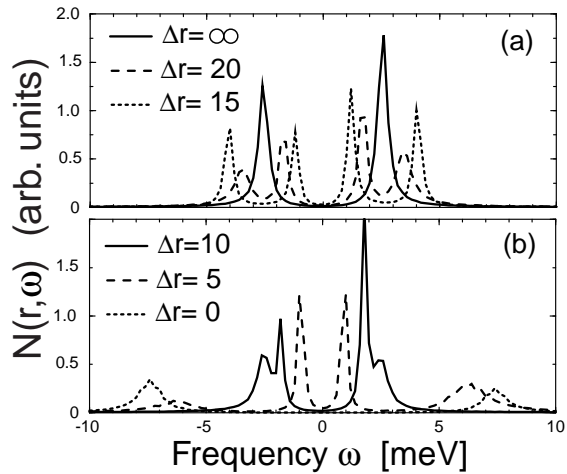}
\end{center}
\caption{Frequency dependence of the chain DOS for two identical
impurities at one of the impurity sites with $\beta = 200$ meV, $U_0 =0$ and
{\it (a)} $\Delta r = \infty$ (solid line), $\Delta r = 20$
(dashed line), and $\Delta r = 15$ (dotted line), and {\it (b)} $\Delta r = 10$
(solid line), $\Delta r = 5$ (dashed line),
and $\Delta r = 0$ (dotted line).} \label{DOS2imp}
\end{figure}

Due to the 1D character of the chains, interference effects
between two impurities should be large and significantly change
the single-impurity DOS discussed above. To study these effects,
we employ a $\hat{T}$-matrix formalism for two identical
impurities where
\begin{eqnarray}
\hat{G}({\bf r},{\bf r}^\prime,i \omega_n) &=& \hat{G}_0({\bf
r},{\bf r}^\prime,i\omega_n) + \nonumber \\ & & \hspace{-2.0cm}
\sum_{n,m=1}^2 \hat{G}_0({\bf r},{\bf r}_m,i\omega_n) \hat{T}({\bf
r}_m, {\bf r}_n, i\omega_n) \hat{G}_0({\bf r}_n,{\bf r}^\prime
,i\omega_n) \ , \label{fullG2}
\end{eqnarray}
$r_{1,2}$ are the positions of the impurities, and
$\hat{T}({\bf r}_m, {\bf r}_n, i\omega_n)$ obeys the
Bethe-Salpeter equation
\begin{eqnarray}
\hat{T}({\bf r}_m, {\bf r}_n, i\omega_n)&=& \hat{V}
\delta_{r_m,r_n} \nonumber \\
& & \hspace{-1cm} + \hat{V} \sum_{l=1}^2 \hat{G}_0({\bf r}_m,{\bf
r}_l,i\omega_n)
\hat{T}({\bf r}_l, {\bf r}_n, i\omega_n)
\ .
\end{eqnarray}
In general, we expect that the chain DOS for two impurities is
similar to that of a single impurity if the inter-impurity
distance, $\Delta r=r_2-r_1$, is larger than the spatial extent of
single impurity  effects, $\xi \approx 10-15$ at $\omega_{res}$ (see
Fig.~\ref{DOSosc}). In contrast, for $\Delta r < \xi$ significant
interference effects from multiple scattering processes should
occur. To test this conjecture, we present in Fig.~\ref{DOS2imp}
the numerically obtained chain DOS at one of the two impurity
sites as a function of $\Delta r$.
For $\Delta r=\infty$ (Fig.~\ref{DOS2imp}a, solid line) we recover
as expected the DOS for a single impurity with two resonance
states at $\omega_{res} =\pm 2.6$ meV (see Fig.~\ref{DOSmag}). As
the distance between the impurities decreases, the interference
effects arising from multiple scattering processes lead to the
formation of bonding and anti-bonding resonances state, and thus
to a splitting of the two resonances into four ($\Delta r=20$,
dashed line). This result provides a possible explanation for the
experimental observation of four resonance states in the chain
DOS \cite{Der01}. The splitting first becomes discernible when
$\Delta r \approx 2\xi$, in agreement with our conjecture above.
The splitting increases until the inter-impurity distance reaches
$\Delta r=15$ (dotted line). For even smaller $\Delta r=10$
(Fig.~\ref{DOS2imp}b, solid line), the peaks recombine again and
spectral weight is transferred to new resonance states at higher
energies ($\Delta r=5$, dashed line). These new states are
precursors of the resonances one obtains in the limit $\Delta r
\rightarrow 0$ (dotted line) when the DOS is that of a single
impurity with twice the scattering strength, $\tilde{\beta}=400$
meV. Note that $\beta=200$ meV is below the critical value,
$\beta_c$, to induce a transition of the ground state, while
$\tilde{\beta}=400$ meV is above this critical value. Thus, the
system can be driven through a quantum phase transition from an
unpolarized to a polarized state by continually varying the
distance between two or more impurities.

In conclusion we have studied the electronic structure near
impurities in the 1D chains of YBa$_2$Cu$_3$O$_{6+x}$. Assuming
that the chains are SC below $T_c$ due to a proximity coupling, we
find that a magnetic impurity induces two resonance states in the
local DOS. The spatial oscillations of these resonances reflect
the particle-hole nature of the chain superconductivity, and are
therefore characteristic features which distinguish it from other
broken symmetry phases. The amplitude of these oscillations
increases below $T_c$, which is consistent with the experimentally
observed broadening of the NQR line width \cite{Gre00}. Moreover,
we argue that the system undergoes a quantum phase transition at a
critical $\beta_c$, where the ground state changes from being
unpolarized for $\beta<\beta_c$ to a polarized one for
$\beta>\beta_c$. Thus, the chain SC state is in many aspects
similar to that of an $s$-wave superconductor. Finally, we show
that  in the presence of a second impurity, interference effects
due to multiple scattering processes lead to a splitting of the
resonances.

We would like to thank J.C. Davis and A. de Lozanne for
stimulating discussions. This work has been supported by DOE at
Los Alamos.

\end{document}